\documentclass[a4paper, doublecol]{epl2}

\usepackage{amsmath}
\usepackage{amsfonts}
\usepackage{amssymb}
\usepackage{upgreek}

\usepackage{subfigure}

\def\a{\upalpha}

\def\be{\begin{equation}}\def\ee{\end{equation}}
\def\bea{\begin{eqnarray}}\def\eea{\end{eqnarray}}
\let \nn  \nonumber

\def\p{\partial}

\def\a{\alpha}

\def\d{\delta}
\def\e{\varepsilon}
\def\g{\gamma}
\def\l{\lambda}
\def\o{\omega}
\def\ph{\varphi}

\title{Single evolution equation describing  nonlinear dynamics \\ of nonlocal optical medium under two-wave mixing}
\shorttitle{Evolution equation for nonlocal dynamical medium}

\author{S.~Bugaychuk\inst{1} \and E.~Tobisch\inst{2} }
\shortauthor{S.~Bugaychuk \& E.~Tobisch }

\institute{
  \inst{1} Institute for Physics,  National Academy of Science, Kiev, Ukraine\\
 \inst{2} Institute for Analysis, Johannes Kepler University, Linz, Austria
}


\abstract{
In this Letter we study theoretically the interaction of optical waves in nonlinear dynamical medium, i.e. medium with relaxation.
Taking into account the relaxation of the photoinduced nonlinearity we
derive a single evolution equation, namely the nonlinear Schr\"{o}dinger equation with coefficients depending on parameters,
for the case of degenerate two-wave mixing at the reflection geometry in bulk Kerr-like medium  possessing a nonlocal nonlinear response.  All coefficients of the single evolution equation for our system are written out explicitly in terms of initial parameters. This is the first  analytical study of the evolution of nonlinear dynamical medium under the action of two wave mixing; usually, it is studied numerically or experimentally making use of some empirical assumptions.
We briefly discuss various possible scenarios of energy transport in the frame of the novel equation.
}

\begin{document}

\maketitle

\section{Introduction}
Nonlinear interaction of coherent laser beams is famous for their self-action processes in optical materials  promising for many applications. The dynamic holography is the most famous in this respect due to its simple realizability in optical communication systems such as all-optical controllable amplifiers, switches, holographic multiplexing, adaptive data storage, optically addressed spatial light modulators, including many-channels systems and optical neural networks. The other huge area is  fiber applications of dynamic gratings for different types of sensors, tunable filters and adaptive interferometers \cite{AllOptical, Holograms2011, Kikuchi, InterferReflect, Fiber2013, GLL2004}. The following three main effects acting simultaneously are the subject of dynamic holography: (1) creation of periodic interference pattern inside a nonlinear medium with the help of two or more laser beams; (2) modulation of the refractive index (and/or the absorbtion coefficient) under action of this interference pattern, or, in other words,  inducing a phase dynamic grating inside a nonlinear medium; (3) self-diffraction of recording beams on the dynamic grating creating by the same beams.

Practical use of waves' diffraction on dynamical gratings  is related to several specific effects which can be implemented depending on the initial parameters of beams coupling. Among these effects are energy and  phase transfer between interacting waves which may occur due to the specific features of nonlinear response in  optical material.

The nonlinear problem of optical two-mixing has been considered thoroughly in the following  way. The coupled-wave equations were derived from the Maxwell's wave equations, see e.g. \cite{Gunter2006,Yeh,Huignard}. There exist two essentially different mathematical approaches used for solving these coupled-wave equations.

In the \emph{first approach},  a general mathematical convention is  used that the  medium response is replaced by the modulation of the refractive index $\Delta n$ and/or by the modulation of the absorption $\Delta \alpha$, so-called the approach of the given gratings. In other words, the medium is regarded as a "black box", without its own intrinsic dynamics, while  effects of the energy transfer and the phase coupling are explained in terms of either shift (the non-local response) or no shift (the local response) of the photoinduced gratings relative to the interference pattern. For instance, in \cite{Huignard}  equations for two-wave mixing in the general form and their solutions have been obtained  in the transmission geometry taking into account many factors, including the photoinduced response of the medium that may be both local and non-local and generates both gratings of the refractive index and of the absorption (or amplification).

In the \emph{second approach}, the study of the nonlinear media takes into account its intrinsic dynamics, including the relaxation of the nonlinearity,
e.g. \cite{BKK,KhyzhBook}. In particular, this approach has been used for detailed characterization of the stationary regime in the transmission four wave-mixing \cite{Bugaich98}. In general,  it allows to  provide explicit description for many characteristics responsible for various nonlinear optical properties of the medium. For instance, explanations can be provided for redistribution of charge carriers in photorefractive crystals and semiconductors; the changes of molecules' orientations in liquid crystals; heat-induced nonlinearity, under given conditions of moving gratings; etc.

The holographic dynamical phase grating is considered as a photoinduced spatial modulation of the refractive index  appearing  during the degenerate (i.e. on the same wave lengths)  wave-mixing  in a Kerr-type nonlinear medium which exhibit a nonlocal response. As a rule, nonlinear mechanism in the dynamical media is regarded in the relation to  diffusion and drift of carriers (responsible for the induced nonlinearity); they occur under the action of time-changing spatially modulated optical field. Thus in one dimensional case, a dynamic grating is formed under the action of a sinusoidal light interference pattern.
If the  grating is varying in time it may be considered as a moving matter-wave of one wavelength described by the $K$ -vector of the grating period.

Furthermore, it has been shown theoretically that in the presence of energy transfer, the amplitude of the grating becomes nonuniform in the medium along the $z$-direction of the light wave propagation \cite{StaSi, HongSaxema}.
The dynamics  of grating can be described by the damped  \emph{sine}-Gordon equation in the transmission geometry or by the damped \emph{tanh}-Gordon equation in the reflection geometry. In the steady state, the grating amplitude admits the form of either a bright-soliton or a dark-soliton solution along the wave-propagation direction \cite{JBH1995, BCPRE2009, BCPRE2012}. The first experimental observation of the localization of the dynamical grating was found in a bulk photo-refractive crystals $LiNbO_3$ during degenerate four-wave mixing \cite{BKMPR}.

In this Letter we regard the dynamical system containing the coupled-wave equations for the two-wave mixing in the reflection geometry, with pure non-local response and neglected absorption. We apply the second approach (see above) and demonstrate for the first time that the complete dynamical system (which includes both the coupled-wave equations and the dynamical equation for the medium) can be  reduced to a single nonlinear evolution equation. This equation  effectively is nonlinear Schr\"{o}dinger equation (NLS) with coefficients depending on parameters. As the solutions of the NLS are fairly well studied in the nonlinear physics, one can operate them for prediction of some  effects in the nonlinear dynamical media in optical system. Especially it concerns the solutions for the transient processes such as regular self-oscillation, periodic doubling and others, as well as their dependence on the parameters of the system.
Main theoretically plausible physical phenomena appearing in the frame of our novel evolution equation are briefly discussed at the end  of the Letter.

\section{Basic mathematical model}
In general, the interaction of waves in optical anisotropic medium is described
by the coupled wave equations  obtained from the Maxwell's equations in the following way.
Total optical field is presented as the sum of all
partial waves taking part in the interaction process, and the dielectric permittivity
 $\e = \e_0 + \Delta \e $ is regarded as a sum of the constant $\e_0$ and a changing part $\Delta \e$.
In the case of two interacting waves with initial amplitudes $\tilde {\bf{E}}_1$,
$\tilde {\bf{E}}_2$ in the Bragg conditions
(when the momentum conversation law is performed as the phase matching condition
${\bf{k}}_1 - {\bf{k}}_2 = {\bf{K}}$, where ${\bf{k}}_i, i=1,2$
is the wave-vector of the $i$-th wave) the coupled wave equations takes the
following form (see e.g. \cite{KhyzhBook, Gunter2006}):
\be
\partial_{\bar{z}} {\bf{E}}_1 = -i \frac{k^2}{2 n k_{1\bar{z}}} \e_1 {\bf{E}}_2, \, \,
\partial_{\bar{z}} {\bf{E}}_2^* = i \frac{k^2}{2 n k_{2\bar{z}}}  \e_1 {\bf{E}}_2^*.
 \label{eq:Eq1}
\ee
Here ${\bf{E}}_i$ is a slow varying amplitude of the $i$-th wave,
$k=k_1 = k_2 = 2 \pi / \l$ is the module of the wave-vector,
$\l$ is the wavelength in a vacuum,
$\bar z [m]$ is the space coordinate in the propagation direction of the waves, and
$t [c]$ is a real time.
The coupling coefficient $\e_1$ in the Eqs.~(\ref{eq:Eq1})
represents the first Fourier component of the
changing part of the dielectric permittivity appearing in an optical medium due
to action of the light field:
$\Delta \e \approx \e_1\exp{\{i K z\} } + c.c.$,
where $K = |{\bf{K}}| = 2 \pi/ \Lambda$, $\Lambda$ is the spatial period
of the light interference pattern. In this way, the photoinduced dielectric
permittivity has the form of the grating with the grating-vector ${\bf{K}}$.

The dynamical nonlinear system also includes the equation describing
the changes of the dielectric permittivity $\Delta \e$.
For the majority of optical nonlinear media this equation may be written by
the following form, \cite{KhyzhBook}:
\be
\frac{\p}{\p t} \Delta \e = \hat {F} (|{\bf{E}}|^2) -
\frac{\Delta \e}{\tau} + D \nabla^2 \Delta \e -
({\bf{\upsilon}}\nabla (\Delta \e)),
\label{eq:Eq2}
\ee
where $\hat {F} (|{\bf{E}}|^2)$, in general case, is an operator depended
on the light intensity $|{\bf{E}}|^2$. For the simplest case of the
cubic nonlinear response (Kerr-like nonlinear media) this operator reads
$\hat {F} (|{\bf{E}}|^2) = \epsilon |{\bf{E}}|^2$.
The other notations in the Eq.~(\ref{eq:Eq2}) are as follows: $D$ is the
diffusion constant, which describes the spatial distribution of the excitations
responsible for the changing dielectric permittivity $\Delta \e$;
$\tau$ is the time relaxation constant, and $\bf{\upsilon}$ is the drift velocity of
these excitations.

Further on we consider the nonlocal media allowing energy transfer
between the interacting waves.
In this case there exists a spatial shift between a local point of the light action and the
nonlinear response of a medium.
For example, in the case of electrooptical crystals with diffusion
mechanism of nonlinearity, the functional $\hat {F} (|{\bf{E}}|^2)$
has the form $\hat {F} (|{\bf{E}}|^2) \simeq ({\bf{C}} \nabla |{\bf{E}}|^2)$,
where ${\bf{C}}$ is the pole axis of the crystal. In this case
$D=0$ and $\tau$ is  finite.

The maximum energy transfer occurs when the spatial shift equals the quarter of the
period of the light interference pattern.
If we present $\e_1$ in terms of the amplitude and the initial phase as
$\e_1 = \tilde \e_1 cos \psi$,
for the case of pure nonlocal response the dielectric permittivity gets the form
$Q = \tilde \e_1 cos(\psi - \pi/2) = -\tilde \e_1 sin(\psi) $
(here notation $Q$ is introduced).

For specific physical medium,  this system may be
complemented by the material equations for nonlinear optical response.
\section{Evolution equations for dynamical nonlinear two-wave mixing}
\begin{figure}
\begin{center}
\includegraphics[width=7.5cm]{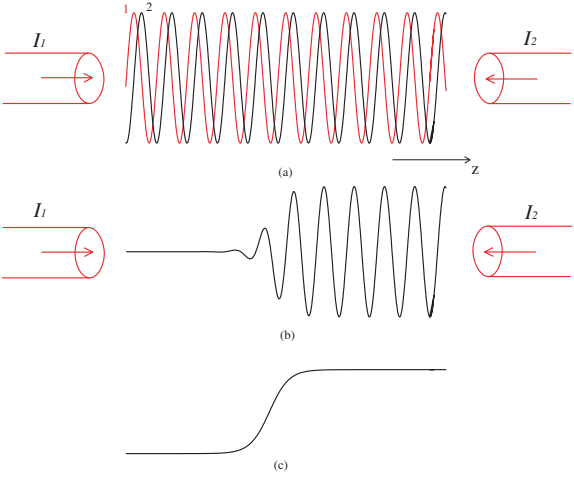}
\end{center}
\caption{\label{f:1} Color online. \textbf{(a)} Formation of the sinusoidal intensity field (the light lattice) by two interacting plane waves $I_1$ and $I_2$ in the reflection geometry - the curve 1.
Modulation of the refractive index (the phase grating) in a nonlocal nonlinear medium, when the refractive index grating is shifted relative to the light
lattice on a quarter of the spatial period - the curve 2.
\textbf{(b)} Change of the grating distribution, when the soliton is formed inside the medium.
\textbf{(c)} The envelope of the grating amplitude in the form of a dark soliton.
}
\end{figure}
We consider the two-wave mixing in the reflection geometry. The case of pure nonlocal response means that
there exists the shift between the interference pattern and the grating on the
quarter of the space period (see Fig.~\ref{f:1}). We also introduce dimensionless coordinate $z, \, z= \bar{z}k^2/2k_{\bar{z}}$, and rewrite
 main coupled-wave equations as follows:
\bea
\p_z \mathbf{E}_1(z,t) = Q(z,t)\mathbf{E}_2(z,t), \, \label{1}\\
\p_z \mathbf{E}_2^*(z,t) = Q(z,t)\mathbf{E}_1^*(z,t)\label{2}.
\eea
These equations describe two slowly changing grating amplitudes of the light beams  $\mathbf{E}_1(z,t)=E_1(z,t) \, \exp{\{i\ph_1\}}$ and $\mathbf{E}_2(z,t)=E_2(z,t) \, \exp{\{i\ph_2\} }$ which are complex-valued functions.

The coupling coefficient $Q(z,t)$  is a real-valued function of two scalar variables given by
\be \label{3}
\p_tQ(z,t)=\g_N \frac{I_m(z,t)}{I_0(z,t)}-\frac{1}{\tau}Q(z,t).
\ee
Here $\g_N$  is the the amplification coefficient of the nonlocal response of the medium correspondingly; it is real. The total intensity $I_0$ and the intensity of the interference pattern $I_m$ are defined by the beams amplitudes as follows:
\bea
I_0&=&I_1(z,t)+I_2(z,t)=|\mathbf{E}_1|^2+|\mathbf{E}_2|^2, \\
I_m&=&(\mathbf{E}_1\mathbf{E}^*_2+\mathbf{E}^*_1 \mathbf{E}_2)=2E_1E_2 \cos{(\ph_1-\ph_2)}.
\eea
Aiming to rewrite Eqs.~(\ref{1})-(\ref{3}) as one equation, we first introduce a new variable $J_m=J_m(z,t)$; it defines modulation depth of the intensity field and is given by the following ratio:
\be
J_m=\frac{I_m}{I_0}=\frac{\mathbf{E}_1\mathbf{E}^*_2+\mathbf{E}^*_1 \mathbf{E}_2}{\mathbf{E}_1\mathbf{E}_1^*+
\mathbf{E}_2\mathbf{E}_2^*}.
\ee

Now let us take derivative over $z$ of the Eq.~(\ref{3}) and rewrite it already in terms of the new variable $J_m$ as
\be \label{4}
\p_t\p_z Q +\frac{1}{\tau}\p_zQ -\g_N \p_z J_m=0.
\ee
Straightforward calculation making use of the Eqs.~(\ref{1}),(\ref{2}) yields
\bea
\p_zJ_m = \p_z \frac{I_m}{I_0}= \frac{I_0 \p_zI_m   - I_m \p_zI_0 }{I_0^2},\\
\p_zI_m = \p_z(\mathbf{E}_1\mathbf{E}_2^*)+\p_z(\mathbf{E}_1^*\mathbf{E}_2),  \\
\p_zI_0 = 2 QI_m, \, \p_z I_m = 2  Q I_0.
\eea
Accordingly, for the  new variable $J_m$ we have
\be \label{5}
\p_z J_m=2(Q-QJ_m^2)
\ee
and after substituting the Eq.~(\ref{5}) into the Eq.~(\ref{4}) we deduce
\bea
\p_{tz}Q+\frac{1}{\tau}\p_{z}Q-2\g_NQ+2\g_NQJ_m^2=0
\eea
which is convenient to rewrite as
\bea \label{eq:Eq6}
\p_{tz}Q+\frac{1}{\tau}\p_{z}Q-2\g_NQ \nn \\
+\frac{2}{\g_N}Q(\p_t Q+ \frac{1}{\tau}Q)(\p_t Q+ \frac{1}{\tau}Q)=0.
\eea
The Eq.~(\ref{eq:Eq6})  is a nonlinear evolutionary PDE of the second order which is equivalent to the initial system of the
Eqs.~(\ref{1})-(\ref{3}).

Our next step is to demonstrate that the Eq.~(\ref{eq:Eq6}) can be  reduced to a parametric  NLS equation.
\section{Reduction to a parametric  NLS}
For our further study it would be convenient to present the function
$Q$ in the following form:
$
Q = \frac{1}{2} \left( {\bf{Q}} + {\bf{Q}}^* \right)
$
where an auxiliary  complex-valued function ${\bf{Q}}$  is introduced
and ${\bf{Q}}^*$ is its complex conjugate.
Further we apply the procedure of the standard perturbation technique, e.g.  \cite{Na81},  to  the Eq.~(\ref{eq:Eq6}),
we expand the auxiliary function ${\bf{Q}}$ into the series
\be
{\bf{Q}} (z,t) = \d {\bf{F}}_0 ,
\label{eq:Eq7}
\ee
but in curved coordinates, assuming that the coordinates $t$ and $z$ are also is expanded into series
\bea
t = T_0 + \d T_1 + \d^2 T_2 + ...\nn ,
\\
z = Z_0 + \d Z_1 + \d^2 Z_2 + ...
\label{eq:Eq8}
\eea
with $\d$ being a small parameter, and  functions ${\bf{F}}_j$ depending on all coordinates,
\be
{\bf{F}}_j = {\bf{F}}_j \left( T_0,T_1,T_2,...,Z_0,Z_1,Z_2,... \right).
\label{eq:Eq9}
\ee
Substituting the series above into the Eq.~(\ref{eq:Eq6}) and
 summing up terms with the same power of the small parameter $\d$ we deduce the Eqs.~(19),(20):
\bea
\left\{
\frac{\p}{\p T_0} \frac{\p}{\p Z_0} +
\d \left[
      \frac{\p}{\p T_0} \frac{\p}{\p Z_1} +
      \frac{\p}{\p T_1} \frac{\p}{\p Z_0}
           \right]\right\} \d {\bf{F}}_0 \nn
\\  +
\left\{ \d^2 \left[
\frac{\p}{\p T_0} \frac{\p}{\p Z_2} +
\frac{\p}{\p T_2} \frac{\p}{\p Z_0} +
\frac{\p}{\p T_1} \frac{\p}{\p Z_1} \right]
\right\} \d {\bf{F}}_0  \times \nn
\\
 +
\frac{1}{\tau} \left\{ \frac{\p}{\p Z_0} + \d \frac{\p}{\p Z_1} +
\d^2 \frac{\p}{\p Z_2} + \d^3 \frac{\p}{\p Z_3} \right\}
\d {\bf{F}}_0  - \nn
\\
\g_N \left[ \d {\bf{F}}_0  \right] +
\frac{2}{\g_N} \hat{R} ({\bf{Q}}) + ... + c.c. =0
 \quad \quad
\Rightarrow
\\
\d \big[ \frac{\p}{\p T_0} \frac{\p}{\p Z_0} {\bf{F}}_0 +
 \frac{1}{\tau} \frac{\p}{\p Z_0} {\bf{F}}_0 - 2\g_N {\bf{F}}_0\big]  + \nn \\
\d^2 \left[
\left( \frac{\p}{\p T_0} \frac{\p}{\p Z_1} +
\frac{\p}{\p T_1} \frac{\p}{\p Z_0} \right) {\bf{F}}_0 +
 \frac{1}{\tau}
\frac{\p}{\p Z_1} {\bf{F}}_0 \right]  + \nn
\\
\d^3
\left[ \frac{\p}{\p T_0} \frac{\p}{\p Z_2} +
\frac{\p}{\p T_2} \frac{\p}{\p Z_0} +
\frac{\p}{\p T_1} \frac{\p}{\p Z_1} \right] {\bf{F}}_0  + \nn
\\
\d^3 \frac{1}{\tau} \frac{\p}{\p Z_2} {\bf{F}}_0
+ \d^3 \frac{2}{\g_N} \hat{R}({\bf{F}}_0)
+ ... + c.c. = 0
\eea

For the brevity of presentation a new notation $\hat{R}({\bf{F}}_0)$ has been introduced
for the last term
\be \mathbf{Q} \left( \p_t \mathbf{Q} + \frac{1}{\tau} \mathbf{Q} \right) \left( \p_t \mathbf{Q} + \frac{1}{\tau} \mathbf{Q} \right)\ee
of the Eq.~(\ref{eq:Eq6}).
\subsection{First order terms,  $\d \bf{F}_0$}
The first order terms generate the equation
\be
 \frac{\p}{\p T_0} \frac{\p}{\p Z_0}
{\bf{F}}_0 + \frac{1}{\tau} \frac{\p}{\p Z_0}{\bf{F}}_0  -
2\g_N {\bf{F}}_0  + c.c.  =0
\ee
which can be rewritten in the operator form as
\be
\hat{L}{\bf{F}}_0 = 0, \, \hat{L} =
\frac{\p}{\p T_0} \frac{\p}{\p Z_0} +
\frac{1}{\tau} \frac{\p}{\p Z_0} -
2\g_N
\label{eq:Eq10}
\ee
where $\hat{L}$ is the operator. Let us look for the solution of
 the Eq.~(\ref{eq:Eq10}) in the form of a two-dimensional plane wave in the coordinates $T_0$ and $Z_0$, but with the amplitude depending  on all \emph{other} coordinates, i.e.
\be \label{e1-1}
{\bf{F}}_0 = {\bf{A}} \left( T_1, T_2, ..., Z_1, Z_2, ...\right) \exp{i \left[ \o T_0 - q Z_0 \right]}
\ee
(physical meaning of $q$  is  grating period). Substituting (\ref{e1-1}) into the Eq.~(\ref{eq:Eq10}) and taking into account that
\be
\frac{\p {\bf{F}}_0}{\p Z_0}= - i q {\bf{F}}_0,\quad
\frac{\p }{\p T_0} \frac{\p {\bf{F}}_0}{\p Z_0} =\o q {\bf{F}}_0,
\ee
it is easy to deduce the form of the dispersion function
\be
\o=i\frac{1}{\tau}+2\g_N\frac{1}{q}
\ee
and the general solution ${\bf{F}}_0$ to the  Eq.~(\ref{eq:Eq10}):
\bea
{\bf{F}}_0 = {\bf{A}} (T_1,T_2,...,Z_1,Z_2,...)  \nn \\
         \times \exp {i \left [ \frac{2\gamma_N}{q} T_0 - q Z_0 + i \frac{1}{\tau} T_0 \right ]}= {\bf{A}} \exp {i {\bf{\Phi}}_0}
\label{eq:Eq12}
\eea
with the complex phase
\be
{\bf{\Phi}}_0 =
2\g_N \frac{1}{q} T_0 - q Z_0 + i \frac{1}{\tau} T_0
\label{eq:Eq13}
\ee
\subsection{Second order terms,  $\d^2 \bf{F}_0$}
These terms generate the following equation on  ${\bf{F}}_0$:
\bea
 \left ( \frac{\p}{\p T_0} \frac{\p}{\p Z_1} +
\frac{\p}{\p T_1} \frac{\p}{\p Z_0} +
\frac{1}{\tau} \frac{\p}{\p Z_1} \right ) {\bf{F}}_0
+ c.c.= 0,
\eea
or in the operator form
\be
\hat{M} {\bf{F}}_0 =0
\label{eq:Eq14}
\ee
where
\be
\hat{M} =  \frac{\p}{\p T_0} \frac{\p}{\p Z_1} +
\frac{\p}{\p T_1} \frac{\p}{\p Z_0} +
\frac{1}{\tau} \frac{\p}{\p Z_1}.
\label{eq:Eq15}
\ee
Substituting the expression (\ref{eq:Eq12}) for ${\bf{F}}_0$ into the Eq.~(\ref{eq:Eq14}) and performing some straightforward calculations of all necessary derivatives one can easily obtain the evolution equation for the amplitude ${\bf{A}}$:
\be
\left (  \frac{\p}{\p T_1} -
\frac{2\g_N }{q^2} \cdot \frac{\p}{\p Z_1} \right )
{\bf{A}} =0.
\label{eq:Eq17}
\ee
It follows that
the variables $T_1$ and $Z_1$ are not
independent and can be replaced by one variable making use of the Galilean transformation
\be
\zeta = T_1 - \upsilon_g Z_1 , \, \, \,
\upsilon_g = \frac{2\g_N}{q^2}
\label{eq:Eq18}
\ee
where the physical meaning of $\upsilon_g $ is the group velocity.
\subsection{Third order terms,  $\d^3 \bf{F}_0$}
Following the same lines as above, we deduce the equation in the operator form
\be
\hat{N} {\bf{F}}_0 +
\frac{2}{\g_N} \hat{R} \left( {\bf{F}}_0 \right) =0
\label{eq:Eq19}
\ee
where the operator $\hat{N}$ reads
\be
\hat{N} = \frac{\p}{\p T_0} \frac{\p}{\p Z_2} +
\frac{\p}{\p T_2} \frac{\p}{\p Z_0} +
\frac{1}{\tau} \frac{\p}{\p Z_2} +
\frac{\p}{\p T_1} \frac{\p}{\p Z_1}.
\ee
Next three steps are: (a) to introduce new variable $\eta= T_2 - \upsilon_g Z_2$ , and
(b) to compute corresponding derivatives
\bea
\frac{\p}{\p \eta} = \frac{\p}{\p T_2}
\frac{\p \eta}{\p T_2} +
\frac{\p}{\p Z_2} \frac{\p \eta}{\p Z_2} =
\frac{\p}{\p T_2} - \upsilon_g \frac{\p}{\p Z_2}\\
\frac{\p}{\p T_1} \frac{\p}{\p Z_1} =
\frac{\p}{\p \zeta} \frac{\p \zeta}{\p T_1}
\left( \frac{\p}{\p \zeta} \frac{\p \zeta}{\p Z_1} \right)  =
 - \upsilon_g \frac{\p^2}{\p \zeta^2}
\eea
and (c) to substitute the results into the Eq.~(\ref{eq:Eq19}):
\be
i \frac{\p {\bf{A}}}{\p \eta} +
\frac{\upsilon_g}{q} \frac{\p {\bf{A}}^2}{\p \zeta^2} -
\frac{2}{q \cdot \g_N}
\exp\{-i {\bf{\Phi}}_0\} \hat{R}  =0.
\label{eq:Eq23}
\ee

Comparing $\hat{R}$
with the Eq.~(\ref{eq:Eq6}) we conclude that
\bea
\hat{R} \left( {\bf{F}}_0 \right) = \frac{1}{2}
\left[ {\bf{F}}_0 + {\bf{F}}^*_0 \right] \times \nn \\
\left[ \partial_t {\bf{F}}_0 + \frac{1}{\tau} {\bf{F}}_0 +
\partial_t {\bf{F}}^*_0 +
\frac{1}{\tau} {\bf{F}}^*_0 \right]^2.
\eea
Thus, we have obtained explicit  dependence of  $\hat{R}$  on  ${\bf{F}}_0={\bf{A}} \exp {i {\bf{\Phi}}_0}$ given by the
Eqs.~(\ref{eq:Eq12}),(\ref{eq:Eq13}). After more calculations (omitted here for the brevity of presentation) we deduce
explicit  expressions for $\hat{R}$
\bea
-\hat{R}/ \frac{8 q^2}{\gamma_N^2}=
{\bf{A}}^3 \exp\{3 i {\bf{\Phi}}_0\} -\nn \\
{\bf{A}} {\bf{A}} {\bf{A}}^* \exp\{i {\bf{\Phi}}_0\}
\exp\{i ( {\bf{\Phi}}_0 - {\bf{\Phi}}^*_0)\}
-\nn \\
{\bf{A}} {\bf{A}}^* {\bf{A}}^* \exp\{-i {\bf{\Phi}}^*_0\}
\exp\{i ( {\bf{\Phi}}_0 - {\bf{\Phi}}^*_0)\} + \nn \\
{\bf{A}}^* {\bf{A}}^* {\bf{A}}^* \exp\{- 3 i {\bf{\Phi}}^*_0\}
\eea
and for
the phase difference
$
{\bf{\Phi}}_0 - {\bf{\Phi}}^*_0 =
2 i \frac{1}{\tau} T_0.
$
Their substitution into the main equation yields
\be \label{eq:Eq25}
i \frac{\p {\bf{A}}}{\p \eta} +
\frac{2\g_N}{q^3} \frac{\p^2 {\bf{A}}}{\p \zeta^2} -
\frac{\g_N}{q^3} \exp{\{-2 T_0/\tau \}
| {\bf{A}} | ^2 {\bf{A}}  } =0,
\ee
or in a more compact form
\be \label{eq:Eq26}
i \frac{\p {\bf{A}}}{\p \eta} +
 2 \a \frac{\p^2 {\bf{A}}}{\p \zeta^2}
- \a \exp{\{-2 T_0/\tau \} }  | {\bf{A}} | ^2 {\bf{A}} =0,
\ee
where  $\a=  q^3 / \g_N$.

This equation describes the evolution of the envelope soliton which is identified by the function $\mathbf{A}$.
The function $\mathbf{A}$ describes  excitation in a nonlinear medium, responsible for the appearance of the nonlinearity. Eventually, the physical realization of this function in real media will be reduced to the envelope of amplitude for photo-induced dynamical grating $\varepsilon_1$. The same soliton-like behavior will be observed for the function $J_m$, which corresponds to the envelope for the intensity values in the interference pattern inside the medium. In our formulation, the output intensities of two interacting waves can easily be found as the functions of the $\mathbf{A}$.

\section{Discussion}\label{s:dis}
As a result of the tedious but straightforward calculations we deduced the evolution equation Eq.~(\ref{eq:Eq26}) having  the form of nonlinear Schr\"{o}dinger equation with coefficients depending on parameters, pNLS.  This allows us to predict various theoretically possible
scenarios of energy transfer in a physical system governed by this evolution equation.

Not going into all details here we mention only a couple of possible choices of parameters
$q, \, T_0, \, \tau, \, \g_N$ yielding completely different energetic behavior of the nonlinear system.  Keeping $\a$ constant, let us just have a look at
the Eq.~(\ref{eq:Eq26}) for different relations between
$\tau$ and $T_0$.

(1) Case  $\tau \ll T_0$. This condition yields than $\a \exp{\{-2 T_0/\tau \} } \approx 0$ turning the Eq.~(\ref{eq:Eq26}) into a linear equation which does not describe nonlinear physical phenomena we are interested in..

(2) Case $\tau \gg T_0.$ Accordingly, we may assume $T_0/\tau \approx 0 \Rightarrow \exp{\{-2 T_0/\tau \} } \approx 1 $ and the Eq.~(\ref{eq:Eq26})  turns into the classical NLS.

In this case the Eq.~(\ref{eq:Eq26}) is integrable by the inverse scattering transform and allows many exact solutions, for instance
\bea\label{solution1}
\bf{A} (\zeta, \eta)= C_1\exp\{i[C_2 \zeta + (\a C_1^2-C_2^2)\eta+ C_3]   \} \quad \mbox{and} \\
 \label{solution2}
\bf{A} (\zeta, \eta) = \frac{C_1}{\sqrt{\eta}} \exp\{ i [\frac{(\zeta+C_2)^2}{4\eta} + \a C_1^2 \ln \eta + C_3] \},
\eea
with $C_i, \, i=1,2,3$ being arbitrary real constants, \cite{PZ2004}. Optical solitons in nonlinear media  with arbitrary dispersion are discussed in \cite{ABA2014} while new general solutions can be found  in \cite{AKCBA2016}.

For identifying another prominent set of solutions, we have to establish whether or not the pNLS is focusing for some combination of parameters, i.e. possesses modulation instability.  This fact is usually established  by the stability analysis, see e.g. \cite{ZakhOstrov2009}, while the answer can be given as following. In optical fibers, the NLS is focusing if the coefficients in front of dispersive and nonlinear terms have different signs; otherwise it is defocusing,  \cite{Agrawal2013}.

Let us compare the signs of the coefficients in front of dispersive and nonlinear terms  in the Eq.~(\ref{eq:Eq26}). The function exponent may only admit nonnegative values; accordingly, regardless of the sign of the parameter $\a$, these coefficients have
opposite signs, i.e. the Eq.~(\ref{eq:Eq26}) is focussing and possesses modulation instability, see e.g. \cite{ZakhOstrov2009}.

 This in turn allows us to conclude that formally  the set of  exact solutions of the Eq.~(\ref{eq:Eq26}) includes also Akhmediev breathers, for instance
\be \label{breather}
\bf{A}_j (\zeta, \eta) = \Big[ (-1)^j + \frac{G_j+iH_j}{D_j}\Big]\exp\{ 2i \zeta\}.
\ee
The coefficients $G_j,\, H_j, \, D_j$ are complex polynomial and their explicit form can be found in \cite{AEK1985}.

Another interesting fact about the pNLS is following: it does not have solutions in the form of dark solitons, as they only appear in the defocusing case, \cite{KL-D1998}.

There exists one more form of energy transport triggered by modulation instability in the pNLS:  formation of dynamical cascades first introduced in \cite{K12a}. Dynamical cascade is a sequence of discrete modes generated by modulation instability under certain conditions allowing to compute the shape of energy spectrum in a nonlinear medium and also to establish some other properties of the cascade (direction, termination, appearing of recurrent FPU-like patterns, etc.), \cite{K12b}.

(3) Case $0< \tau / T_0 < \infty$. This means that the ratio is finite and that the coefficient  $\a \exp{\{-2 T_0/\tau \} }$ located in front of the nonlinear term of the pNLS is decaying exponentially thus allowing to regard the evolution equation  (\ref{eq:Eq26}) as a parametrical-damped NLS (pdNLS), for this choice of parameters.

Unlike the previous, well-studied cases, this case seems to be the most interesting and the less studied one;
there is no consensus in the scientific literature concerning this case. Some researchers argue that any arbitrarily small dissipation suppresses modulation instability \cite{StabilizingMI2005}. Other researchers  demonstrate that dissipation indeed affects the modulation
instability, e.g. alters the growth rates of sideband perturbations and the boundaries of the linearized stability domains in the modulation wavevector space. Moreover, the instability occurs only if the amplitude of the background wave exceeds a certain
threshold as it is demonstrated in \cite{VShTh2008},  where the study has been conducted for the NLS applied for water waves. In the context of nonlinear optics these effects would be convenient to discussed in terms of dissipative solitons, \cite{AA2005}, which are not energy conserving, in contrast to classical solitons mentioned above. This conception provides "an excellent framework for understanding complex pulse dynamics and stimulates innovative cavity designs", \cite{GA2012}.

Last but not the least. In contrast to the modulation instability, resonant interactions are possible for arbitrary small non-zero initially excited  $\bf{A}$, though time scales and/or characteristic times of interaction would differ from those of modulation instability (see \cite{K13} for more details). The most important property of the resonant interactions is that they  may form independent clusters of resonant triplets of quartets such that waves in a cluster exchange energy only among themselves; no energy is leaving a cluster, at the corresponding time scale. Detailed description of fast computing procedures for the construction of such clusters in various physical systems are described in \cite{KK06-1,KK06-2,KM07}.%

Structure of resonance clusters appearing in the NLS (exemplified with water waves) is discussed e.g. in \cite{K07, KNR08} while examples of their dynamical behavior in Hamiltonian systems has been studied in \cite{BK09_1, BK09_2}. Energy exchange within a resonant cluster may be periodic or chaotic, depending on many parameters of its dynamical system and also on the initial energy distribution, \cite{CUP}.  Statistical description of a big number of connected resonance  clusters in nonlinear optics (so-called optical wave turbulence) is given in
\cite{PGHSRMC2014}.

The identification and interplay  of all possible phenomena mentioned above for physically relevant set of parameters and time scales is presently under the study.
\section{Conclusions}
(1)  We have deduced a single evolution equation called parametric NLS (pNLS) describing time evolution of bulk Kerr-like medium  possessing a nonlocal nonlinear response under the action of the degenerate two-wave mixing at the reflection geometry.
The same equation can be deduced for the degenerate four-wave mixing,  two-wave system has been taken for the sake of simplicity of presentation.
The main novelty in our approach is that we regard the nonlinear dynamical medium, i.e. the relaxation of the   photoinduced nonlinearity is not constant, in contrast to the common approach widely developed in nonlinear optics.

(2) We briefly presented a few theoretically possible scenarios of energy transport in the frame of the pNLS (exact solutions, dynamical cascades, dissipative solitons, resonance clusters, optical wave turbulence). Whether or not all these phenomena can be observed in practice will depend on the combination of laser and medium parameters available in an experiment.

(3) The developed theoretical findings can be experimentally realized  in any photorefractive medium, either in
traditional photorefractive crystals, liquid crystals, media with thermal or saturable absorption
nonlinearity, or in modern photonic crystals, semiconductor heterostructures and nanocomposites with quantum dots that possess ultrafast nonlinear response.

It follows from our theoretical model that there exist an intricate link between the phase grating of a matter-wave and the light lattice formed by actively interacting waves. Emergence of the nonlocal response in a medium is identified by the appearance of energy transfer between interacting waves; this in turn implies
that the grid-generated soliton is formed inside the medium. Parameters of this soliton depend explicitly on the characteristics
both of the photorefractive medium and of the input waves (their mutual intensities, phases and polarization states),
and can be used for forecasting of various effects appearing due to the two-wave mixing, see e.g.
\cite{BCPRE2012}.
An experimental realization of individual scenarios of
controllable energy transfer in nonlinear liquid crystal cells
is a subject for our further study.

\textbf{Acknowledgment} E.T. acknowledges the
support of the Austrian Science Foundation (FWF) under project
P24671. The authors are very grateful to Shalva Aminashvili for fruitful discussions and to the both Reviewers for their valuable comments which allowed significantly improve the presentation of our results.

 \end{document}